\documentclass[aps,prd,amsmath,showpacs,showkeys,preprint]{revtex4}
\usepackage{amsmath}
\usepackage[cp1251,koi8-r]{inputenc}
\usepackage[english]{babel}
\usepackage{epsfig}
\usepackage{graphics}

\newcommand{\be}{\begin{equation}}
\newcommand{\ee}{\end{equation}}
\newcommand{\bn}{\begin{eqnarray}}
\newcommand{\en}{\end{eqnarray}}
\newcommand{\bd}{\begin{displaymath}}
\newcommand{\ed}{\end{displaymath}}
\newcommand{\bs}{\begin{subequations}}
\newcommand{\es}{\end{subequations}}
\newcommand{\bsp}{\begin{split}}
\newcommand{\esp}{\end{split}}
\newcommand{\bnn}{\begin{eqnarray*}}
\newcommand{\enn}{\end{eqnarray*}}
\newcommand{\ba}{\begin{aligned}}
\newcommand{\ea}{\end{aligned}}

\def\Journal#1#2#3#4#5#6{#1, \ #2, \ #3 \ #4 \ (#5) \ #6.}

\begin{document}

\inputencoding{cp1251}

\title{Cross-correlation markers \\ in stochastic dynamics of complex systems}

\author{\firstname{O.~Yu.}~\surname{Panischev}}
\email{opanischev@gmail.com}
\affiliation{Department of Physics, Kazan State University,
420008 Kazan, Kremlevskaya Street, 18 Russia}
\author{\firstname{S.~A.}~\surname{Demin}}
\email{serge_demin@mail.ru}
\affiliation{Department of Physics, Kazan State University,
420008 Kazan, Kremlevskaya Street, 18 Russia}
\author{\firstname{J.}~\surname{Bhattacharya}}
\affiliation{Department of Psychology, Goldsmiths College, University of London, New Cross, London SE14 6NW, UK}
\affiliation{Commission for Scientific Visualization, Austrian Academy of Sciences, Vienna, A1220, Austria}

\begin{abstract}

The neuromagnetic activity (magnetoencephalogram, MEG) from healthy human brain and from an
epileptic patient against chromatic flickering stimuli
has been earlier analyzed on the basis
of a memory functions formalism (MFF). Information measures
of memory as well as relaxation parameters revealed high individuality and unique features in
the neuromagnetic brain responses of each subject. The current paper demonstrates new
capabilities of MFF by studying cross-correlations between MEG signals obtained from
multiple and distant brain regions. It is shown that the MEG signals of healthy subjects
are characterized by well-defined effects of frequency synchronization and at the same time by the
domination of low-frequency processes. On the contrary, the MEG of a patient is characterized by a
sharp abnormality of frequency synchronization,
and also by prevalence of high-frequency quasi-periodic processes.
Modification of synchronization effects and dynamics of
cross-correlations offer a promising method of detecting pathological abnormalities in
brain responses.
\end{abstract}



\maketitle

\section{Introduction. Synchronization and collective effects in time series analysis of complex systems}

One of the main factors determining the evolution of
complex systems is the presence of collective effects
arising from an interacting or
redistributing of the certain connections between parts
of composite system. In many cases it is impossible to
make an adequate analysis of the functions of the systems by
ignoring the underlying collaborative mechanisms.

There are various approaches used in studying the collective
phenomena in complex systems. Somehow or other, all of them are
based on the analyzing unique features of the connected
systems: certain quantitative and qualitative ratios between the
system elements, a dynamic coordination of components under the
external influences, specific synchronization phenomena. Some recent results
have been derived by studying the
effects of frequency and phase synchronization
\cite{Gilb,Morm1,Quan,phase,Wavel}. These methods are based on
revealing the characteristic frequencies and analyzing the
differences in the phases of dynamic variables derived by means of
the Fourier transform, Hilbert transform \cite{Gilb,Morm1,phase}
and wavelet-transform \cite{Wavel}. Within the framework of
another methodology the stochastic synchronization is studied by
comparing topological structures of attractors, describing the
dynamics of two nonlinear coupled oscillators \cite{Stoch}. The
``generalized synchronization'' relationship \cite{Rulk} also
uses the topological method and is the successful original
development of the stochastic synchronization approach.

Another approach to study
the collective effects in complex systems is the analysis of cross-correlations,
i.e. the probabilistic relation between the
sequences of random variables. The cross-correlation method is
used to describe the collective phenomena in various systems
(physical, economical, biological and physiological). The perspective approach in this field of research is detrended cross-correlation analysis, which has been introduced in \cite{Poprl} to study the power-law cross-correlations between nonstationary time series of various nature. This method is widely used in financial systems \cite{Zhou,Popnas,Poeur}.

There are several alternative methods to analyze cross-correlations, particularly, random matrix theory \cite{Pler}, approximate entropy \cite{Pinc} and sample entropy \cite{Moor}. In \cite{Pler}, the authors used random matrix theory, a method originally developed to study the spectra of complex nuclei, to analyze the mutual dynamics in price changes of the stocks. In \cite{Pinc}, the authors used the method of approximate entropy, a model independent measure of sequential irregularity which is based on Kolmogorov entropy, as an indicator of system stability. In \cite{Moor}, the authors proposed the sample entropy, a modified and unbiased version of approximate entropy, as a measure of degree of asynchrony in physiological signals.

Although these earlier methods find useful applications in real-life time series, there exists another set of methods
in cross-correlation analysis which uses the cross-correlation functions itself \cite{Quan,Ginz,Long}. In paper \cite{Quan}, the authors compared the linear synchronization measures including the cross-correlation functions to nonlinear ones and revealed that for the considered
experimental data all measures ranked the synchronization levels of the three examples in the same way.
In work \cite{Ginz}, the authors develop a theory of neuronal cross-correlation functions
for analyzing the neuron interactions in large neural networks
including several highly connected sub-populations of neurons.
In work \cite{Long},  the authors developed a method based on the
model of fractional Brownian motion and Hurst exponent to describe the
coupling of the non-stationary signals with long-range correlations.

In this study, we demonstrate the fundamentally new opportunities of
studying the collective effects in complex systems. Our method is
based on generalization of the memory functions formalism
\cite{rmy1,rmy2} in a case of cross-correlations between the
spaced elements of the studied system. The important advantage of
this method is the description of the cross-correlations in different relaxation scales
in the time series of a complex systems. Here, we consider neuromagentic
responses (magnetoencephalogram, MEG) of brain as a suitable example of a time series of a complex system, i.e.
the human brain. Earlier analysis of MEG signals \cite{rmy3} were performed on the basis of the memory
functions formalism (MFF), and revealed an important autocorrelation
difference between MEG signals of healthy subjects and of a patient
with photosensitive epilepsy (PSE).
Particularly, this difference appears in qualitative alterations of the power
spectra of memory functions. Besides, it has been shown that the
statistical memory effects play a key role in identification of
PSE.

Here we investigate cross-correlations in the MEG responses simultaneously obtained from
multiple brain regions. We will show that mechanisms of  formation of the PSE are
connected, first of all, with abnormality of interrelations
between the spaced areas of a cerebral cortex, which result in
suppression of its regulator functions at formation of the
response to external influences. The paper is structured as follows.
Section 2 presents the basic relations of the memory
functions formalism for cross-correlations. Section 3 details the experimental details.
Section 4 contains our results including the calculation of cross-
correlation functions, memory functions and their power spectra.
Section 5 offers the general conclusions about connections between the PSE
pathological alterations and suppression of coordination effects
(frequency-phase synchronization).

\section{Basic relations of the memory functions formalism in case of cross-correlations}
Following \cite{rmy1,rmy2,rmy3,Gasp}, we consider stochastic
dynamics of the magnetic induction gradient, registered in two
different brain regions as the sequences $\{x_j\}, \{y_j\}$ of random
values $X, Y$: \be\ba X=&\{x(T), x(T+\tau), x(T+2\tau), \ldots,
x(T+(N-1)\tau)\},\\
Y=&\{y(T), y(T+\tau), y(T+2\tau), \ldots, y(T+(N-1)\tau)\}
\label{ts},\ea \ee where $T$ is the initial time point,
$(N-1)\tau$ is the time period of signal registration, $\tau$ is the time interval of signal discretisation.

Mean values, fluctuations and dispersions for a set of random values
\eqref{ts} can be written as follows \be \ba \nonumber
\langle X \rangle = \frac{1}{N} \sum_{j=0}^{N-1} x(T+j\tau),
\qquad x_j=x(T+j\tau),& \qquad \delta x_j=x_j-\langle X \rangle,
\qquad \sigma^2_x=\frac{1}{N} \sum_{j=0}^{N-1} \delta x_j^2; \\
\nonumber \langle Y \rangle = \frac{1}{N} \sum_{j=0}^{N-1}
y(T+j\tau), \qquad y_j=y(T+j\tau),& \qquad \delta y_j=y_j-\langle
Y \rangle, \qquad \sigma^2_y=\frac{1}{N} \sum_{j=0}^{N-1} \delta
y_j^2. \ea \ee

To describe the probabilistic relation between the sequences of
random variables $X$ and $Y$ we use
the normalized time-dependent cross-correlation function (CCF): \be
c(t)=\frac{1}{(N-m)\sigma_x\sigma_y}\sum_{j=0}^{N-m-1}\delta
x(T+j\tau)\delta y(T+(j+m)\tau),\label{CCF} \ee \be \nonumber
t=m\tau, \qquad 1\leq m \leq N-1. \ee Function \eqref{CCF} satisfies the conditions
of normalization and relaxation of correlations: \be \nonumber \lim_{t \to
0}c(t)=1, \qquad \lim_{t \to \infty}c(t)=0. \ee It should be noted that the second property
not always satisfied for the real time series of complex systems.

Taking into account the fact that the process is discrete, we must rearrange
all standard operation of differentiation and integration
\cite{diskr1,diskr2}:
\be \ba \frac{dx}{dt}&\rightarrow \frac{\Delta x(t)}{\Delta t} = \frac{x(t+\tau)-x(t)}{\tau}, \\
\int_a^b x(t)dt=\sum_{j=0}^{n-1} x(T_a+j\tau)\Delta
t=~&\tau\sum_{j=0}^{n-1} x(T_a+j\tau)=n\tau\langle X \rangle, \
b-a=c, c=n.\tau \ea \label{devint}\ee

Following \cite{rmy1,rmy2,rmy3}, to describe the dynamics of
studied variables let's introduce the evolution operator $U(t+\tau,
t)$: \be\ba
x(t+\tau)=U(t+\tau,t)x(t),\\
y(t+\tau)=U(t+\tau,t)y(t). \label{evo3} \ea\ee Using Eqs.
\eqref{devint} and \eqref{evo3}, we can establish a formal
equation of motion for any $x_j$ and $y_j$: \be\ba
\frac{\Delta x_j(t)}{\Delta t}=\frac{x_{j+1}(t+\tau)-x_j(t)}{\tau}=\tau^{-1}\{U(t+\tau,t)-1\}x_j(t),\\
\frac{\Delta y_j(t)}{\Delta
t}=\frac{y_{j+1}(t+\tau)-y_j(t)}{\tau}=\tau^{-1}\{U(t+\tau,t)-1\}y_j(t)
\ea\ee and then introduce a Liouville's quasioperator $\hat{L}$ as follows:
\be\ba
\frac{dx(t)}{dt} &\rightarrow \frac{\Delta x(t)}{\Delta t}=i\hat{L}(t, \tau)x(t),\\
\frac{dy(t)}{dt} &\rightarrow \frac{\Delta y(t)}{\Delta t}=i\hat{L}(t, \tau)y(t),\\
\hat{L}(t, \tau)&=(i\tau)^{-1}[U(t+\tau, t)-1].\label{Liuv1}
\ea\ee

Analogously by \cite{rmy1,rmy2,rmy3,Gasp}, let's present the sets of fluctuations
$\delta x_j=\delta x(T+j\tau),$ $\delta y_j=\delta y(T+j\tau),$
where $j=0, 1, \ldots, N-1,$ as the $k$-component vectors
of system state in Euclidean space of the vectors of state:
\be \ba &\textbf{A}_k^0=\textbf{A}_k^0(0)=\{\delta
x_0,\delta x_1, \ldots,\delta x_{k-1}\}, \\
&\textbf{B}_k^0=\textbf{B}_k^0(0)=\{\delta y_0,\delta y_1,
\ldots,\delta y_{k-1}\}. \label{ab0} \ea \ee $k$-component
vectors of system state in time moment $t$:
$\textbf{A}_{m+k}^m(t)$, $\textbf{B}_{m+k}^m(t)$ are arose by the
time shift on step $t=m\tau$: \be \ba
\textbf{A}_{m+k}^m=\textbf{A}_{m+k}^m(t)=&\{\delta x_m,\delta
x_{m+1}, \ldots,\delta x_{m+k-1}\}, \\
\textbf{B}_{m+k}^m=\textbf{B}_{m+k}^m(t)=&\{\delta y_m,\delta
y_{m+1}, \ldots, \delta y_{m+k-1}\}. \label{abt} \ea \ee These
vectors can be derived by multifold actions of the evolution
operator $U(t+\tau,t)$ to vectors of the initial system state: \bs
\bn \textbf{A}_{m+k}^m(t)=U(T+m\tau,T)\textbf{A}_k^0(0),
\label{au} \\ \textbf{B}_{m+k}^m(t)=U(T+m\tau,T)\textbf{B}_k^0(0)
\label{bu}. \label{bt} \en \es

Further we will introduce the scalar product operation: \be
\nonumber \langle \textbf{A}_k^0 \ \textbf{B}_{m+k}^m
\rangle=\sum_{i=1}^k A_i^0 \ B_{m+i}^m.\  \ee

Using the Eqs. \eqref{CCF} -- \eqref{bu}, we can rewrite equation
for CCF as a scalar product of vectors $\textbf{A}_k^0(0)$ and
$\textbf{B}_{m+k}^m(t)$: \be c(t)=\frac{\langle \textbf{A}_k^0(0) \
U(T+m\tau,T) \ \textbf{B}_k^0(0)\rangle}{\langle \textbf{A}_k^0(0) \
\textbf{B}_k^0(0) \rangle}=\frac{\langle \textbf{A}_k^0(0) \
\textbf{B}_{m+k}^m (t) \rangle}{\langle \textbf{A}_k^0 (0) \
\textbf{B}_k^0 (0) \rangle}. \label{CCF2} \ee

Let's write the finite-difference Liouville's equation for the state vectors
(see \eqref{Liuv1}): \be
\frac{\Delta}{\Delta t} \ \textbf{B}_{m+k}^m (t)=i\hat{L}(t,\tau)
\ \textbf{B}_{m+k}^m (t). \label{Liuv2} \ee

Let's introduce the operators of projection $\Pi$ and $\mathrm P$ for the Euclidean
space of state vectors: \be \Pi=\frac{|\textbf{B}_k^0
(0)\rangle \langle \textbf{A}_k^0 (0)|}{\langle\textbf{A}_k^0(0) \
\textbf{B}_k^0 (0)\rangle}, \ \Pi^2=\Pi, \ \mathrm P=1-\Pi, \
\mathrm P^2=\mathrm P, \ \Pi\mathrm P=\mathrm P\Pi=0.\ee

The initial CCF \eqref{CCF2} can be derived by projection of state
 vector $\textbf{B}_{m+k}^m (t)$ to vector of the initial system state
$\textbf{B}_k^0 (0)$: \be \nonumber \Pi \textbf{B}_{m+k}^m (t)
\rangle=\textbf{B}_k^0 (0) \rangle \frac{\langle \textbf{A}_k^0(0)
\ \textbf{B}_{m+k}^m (t) \rangle}{\langle \textbf{A}_k^0 (0) \
\textbf{B}_k^0 (0) \rangle}= \textbf{B}_k^0(0)\rangle c(t).\ee
Projection operators $\Pi$ and $\mathrm P$ split the Euclidean vector
space into two mutually-orthogonal subspaces: \be \nonumber
\textbf{B}(k)=\textbf{B}'(k)+\textbf{B}''(k), \
\textbf{B}'(k)=\Pi\textbf{B}(k), \ \textbf{B}''(k)=\mathrm P
\textbf{B}(k), \ \textbf{B}_{m+k}^m(t) \in \textbf{B}(k).\ee
This permits to split Liouville's equation \eqref{Liuv2} into
two equations within two mutually supplementary subspaces as follows:
\bs \bn \frac{\Delta\textbf{B}'(t)}{\Delta
t}=i\hat{L}_{11}\textbf{B}'(t)+i\hat{L}_{12}\textbf{B}''(t),
\label{Lv_1}
\\ \frac{\Delta\textbf{B}''(t)}{\Delta
t}=i\hat{L}_{21}\textbf{B}'(t)+i\hat{L}_{22}\textbf{B}''(t).
\label{Lv_2}
\en \es Here $\hat{L}_{ij}=\Pi_i \hat{L}\Pi_j, \ i,j=1,2, \ \Pi_1=\Pi,
 \ \Pi_2=\mathrm P,$ are the matrix elements of Liouville's quasioperator: \be \nonumber
\hat{L}=\hat{L}_{11}+\hat{L}_{12}+\hat{L}_{21}+\hat{L}_{22}.\ee
To transit between the subspaces the operators $\hat{L}_{ij}$ are used in the following way: \be \nonumber
\hat{L}_{11} \text{--- from } \textbf{B}' \text{ to } \textbf{B}', \quad \hat{L}_{12}
\text{--- from } \textbf{B}'' \text{ to } \textbf{B}', \quad \hat{L}_{21}
\text{--- from } \textbf{B}' \text{ to } \textbf{B}'', \quad
\hat{L}_{22} \text{ --- from } \textbf{B}'' \text{ to } \textbf{B}''.
\ee

Solving the Eq. \eqref{Lv_2} and using the derived results in Eq.
\eqref{Lv_1} we come to the closed finite-difference discrete
equation for the initial CCF (se for details \cite{rmy1}): \be
\frac{\Delta c(t)}{\Delta
t}=\lambda_1^{XY}c(t)-\tau\Lambda_1^{XY}\sum_{j=0}^{m-1}
M_1^{XY}(j\tau)c(t-j\tau). \label{eq0} \ee Here $\lambda_1^{XY}$
is an eigen
 frequency of Liouville's quasioperator $\hat{L}$, $\Lambda_1^{XY}$ is a
relaxation parameter with the dimension of squared frequency, $M_1^{XY}(j\tau)$ is a normalized cross-correlation memory function of the first order: \be \nonumber
\lambda_1^{XY}=\frac{\langle\textbf{A}_k^0(0)\hat{L}\textbf{B}
_k^0(0)\rangle}{\langle\textbf{A}_k^0(0)\textbf{B}_k^0(0)\rangle},
\quad
\Lambda_1^{XY}=\frac{\langle\textbf{A}_k^0(0)\hat{L}^2\textbf{B}
_k^0(0)\rangle}{\langle\textbf{A}_k^0(0)\textbf{B}_k^0(0)\rangle},
\ee \be M_1^{XY}(j\tau)=\frac{\langle
\textbf{A}_k^0(0)\hat{L}_{12}\{1+i\tau\hat{L}_{22}\}\hat{L}_{21}
\textbf{B}_k^0(0)\rangle}{\langle\textbf{A}_k^0(0)\textbf{B}_k^0(0)
\rangle}, \quad M_1^{XY}(0)=1. \label{M1}\ee

Using Eqs. \eqref{CCF2} -- \eqref{eq0}, we can introduce a set of
projection operators $\Pi_n$ and $\mathrm P_n$ and derive a chain
of equations for cross-correlation memory functions of the $n-1$
order: \be \frac{\Delta M_{n-1}^{XY}(t)}{\Delta
t}=\lambda_n^{XY}M_{n-1}^{XY}(t)-\tau\Lambda_n^{XY}\sum_{j=0}^{m-1}
M_n^{XY}(j\tau)M_{n-1}^{XY}(t-j\tau). \label{eq2}\ee To derive the
kinetic and relaxation parameters and also memory functions we
will use the Gram-Schmidt orthogonalization procedure: $\langle
\textbf{W}_n^X \textbf{W}_m^Y\rangle=\delta_{n,m}\langle
\textbf{W}_n^X \textbf{W}_n^Y\rangle$, where $\delta_{n,m}$ is
Kronecker's symbol. Using it we may easily introduce the
recurrence formula for the dynamic orthogonal variables
$\textbf{W}_n^X$, $\textbf{W}_n^Y$ in which the senior values are
connected with the junior values: \bnn \textbf{W}_0^X
=\textbf{A}_k^0(0), \
\textbf{W}_1^X=(i\hat{L}-\lambda_1^{XY})\textbf{W}_0^X, \
\textbf{W}_2^X=(i\hat{L}-\lambda_2^{XY})\textbf{W}_1^X-\Lambda_1^{XY}
\textbf{W}_0^X-\ldots \ , \\ \textbf{W}_0^Y =\textbf{B}_k^0(0), \
\textbf{W}_1^Y=(i\hat{L}-\lambda_1^{XY})\textbf{W}_0^Y, \
\textbf{W}_2^Y=(i\hat{L}-\lambda_2^{XY})\textbf{W}_1^Y-\Lambda_1^{XY}
\textbf{W}_0^Y-\ldots \ . \enn

Then the eigenvalues of Liouville's quasioperator $\lambda_n^{XY}$ and relaxation parameters $\Lambda_n^{XY}$
in Eq. \eqref{eq2} will be: \be \nonumber
\lambda_n^{XY}=\frac{\langle\textbf{W}_{n-1}^X\hat{L}\textbf{W}_{n-1}^Y
\rangle}{\langle\textbf{W}_{n-1}^X\textbf{W}_{n-1}^Y\rangle}, \
\Lambda_n^{XY}=i\frac{\langle\textbf{W}_n^X\textbf{W}_n^Y
\rangle}{\langle\textbf{W}_{n-1}^X\textbf{W}_{n-1}^Y\rangle}.\ee
Normalized cross-correlation memory function in \eqref{eq2} will be: \be\nonumber
M_{n-1}^{XY}(t)=\frac{\langle\textbf{W}_{n-1}^X\{1+i\tau\hat{L}_{22}\}^m\textbf{W}_{n-1}^Y
\rangle}{\langle\textbf{W}_n^X\textbf{W}_n^Y\rangle}.\ee

A relaxation time of the initial CCF and memory functions of the $n$ order
are determined as follows: \be \tau_c=\Delta
t\sum_{j=0}^{N-1}c(t_j), \ \ldots \ , \
\tau_{M_n^{XY}}=\Delta t\sum_{j=0}^{N-1} M_n^{XY}(t_j). \ee
The set of dimensionless values will determine the statistical spectrum of non-Markovian parameter,
the informational measure of memory:
\be
\left\{\varepsilon_i^{XY}\right\}=\left\{\varepsilon_1^{XY},\varepsilon_2^{XY},
\ldots \ , \varepsilon_{n-1}^{XY}\right\}, \quad
\varepsilon_1^{XY}=\frac{\tau_c}{\tau_{M_1^{XY}}}, \
\ldots, \ \varepsilon_{n-1}^{XY}=\frac{\tau_{M_{n-1}^{XY}}}{\tau_{M_n^{XY}}}.
\ee Thus the value $\varepsilon_{n-1}^{XY}$ is an useful criterium for comparison of
relaxation times of the memory functions $M_{n-1}^{XY}$ and $M_n^{XY}$. This quantitative criterion
characterizes quantitatively the degree of Markovity of the processes and the memory effects in discrete dynamics of complex systems.

In this paper we use the frequency-dependent case of information memory measure:
\be
\varepsilon_i^{XY}=\left\{
\frac{\mu_{i-1}^{XY}(\nu)}{\mu_i^{XY}(\nu)}\right\}^{\frac{1}{2}}.\ee
Here $\mu_i^{XY}(\nu)$ is a power spectrum of $i$th memory function: \be\nonumber \mu_0^{XY}(\nu)=\left | \Delta t
\sum_{j=0}^{N-1}c(t_j)\cos 2\pi\nu t_j\right |^2, \
\mu_1^{XY}(\nu)=\left | \Delta t \sum_{j=0}^{N-1}M_1^{XY}(t_j)\cos
2\pi\nu t_j\right |^2, \ldots \ , \ee \be \nonumber
\mu_i^{XY}(\nu)=\left | \Delta t \sum_{j=0}^{N-1}M_i^{XY}(t_j)\cos
2\pi\nu t_j\right |^2. \ee

The preliminary analysis of experimental series of complex systems
shows, that the special interest for analyzing the statistical
memory  effects represents the ultralow frequency area of the
information characteristic $\varepsilon_1^{XY}(\nu \rightarrow
0)$: \be
\varepsilon_1^{XY}(0)=\left\{\frac{\mu_0^{XY}(0)}{\mu_1^{XY}(0)}\right\}^{\frac{1}{2}}.
\ee This area determines the long range correlations in
experimental series $\{x_j\}$, $\{y_j\}$.

\section{Registration of neuromagnetic activity in human cerebral cortex}

The analyzed MEG signals \cite{PSE1,PSE2} are the induced
neuromagnetic responses obtained from a group of neurologically healthy participants and of a
patient diagnosed with PSE while they were viewing flickering stimuli
with various chromatic combintions.
The PSE is a form of reflexive epilepsy in which the seizures are
provoked by various forms of visual stimuli. Amongst various parameters of a visual stimulus,
chromaticity is less studied in the context of PSE, yet flickering colourful stimuli
are quite widespread in modern technical age laden with multimedia gadgets.
In previous works \cite{PSE1,PSE2}, the authors used the whole-head
MEG system (\emph{Neuromag-122, Neuromag Ltd. Finland}) by means
of 61 SQUID-sensors (superconducting quantum interference device)
and recorded neuromagnetic brain responses from nine healthy or control subjects
(age range 22--27 yrs) and a patient with PSE (age 12 yr). Subjects
of control group had no personal or family history of
photosensitive epilepsy.  All subjects were explicitly informed
that flicker stimulation might lead to epileptic seizures. In order to limit the
health risk, stimuli with only short duration (2 s) was used. All participants
gave their written informed consent before recording. Adequate clinical
protocol was maintained during the recording from the patient.
The subjects were instructed to passively observe visual stimuli with minimal
eye movement. Visual stimuli (red-blue, red-green) were generated
by using two video projectors (Sharp XV-E500, Japan), each of
which produced a continuous single color. For each trial, the
flickering stimulus was presented for 2 s; the gap between trials
was 3 s. Evoked MEG responses were obtained by traditional
averaging technique across artifact-free trials; at least 80
artifact-free trials were averaged for each stimulus. The sampling
frequency was 500 Hz.

To study cross-correlations from various combinations of the MEG signals, we
have chosen a subset of sensors from selected brain regions as follows: occipital (sensor no. 51, 52, 53),
left-temporal (No. 30, 34) and right-temporal (No. 56, 57). Our choice is determined by the
physiological mechanism of processing and transferring the visual
information in a brain. The visual cortex, in which the signals
are transferred from the retina via thalamus, is in occipital area.
Temporal areas play an important role in the perception of the
information, including visual one. Out of three chromatic flickering
stimuli (Red-Blue, Blue-Green, and Red-Green), we have chosen to analyze
the MEG signals against Red-Blue flickering stimulus only, as this was shown
to cause the largest neuromagnetic responses \cite{PSE1,PSE2}. For the generalized description of dynamics of the cross-correlations
we consider the sixth subject from control group.

\section{Cross-correlations in induced neuromagnetic activity of human brain}

\subsection{The PSE-induced stratification of phase clouds of the dynamic orthogonal variables}

Figs. 1, 2 contain the initial time series registered by the 56th
(right-temporal area) and the 51st (occipital area) SQUIDs of the
healthy subject (Fig. 1) and the patient with PSE (Fig. 2). MEG
dynamics of healthy subject is characterized by the significant
large-scale fluctuations while in patient the small-scale
fluctuations are manifested against the background of
quasi-periodic oscillations. Quasi-periodic structure of the
signal in Fig. 1 is directly connected with the physiological
rhythms of electromagnetic brain activity in healthy subject. In
the first 200 ms the flickering stimulus is switched off, the
control signal is registered. It is characterized by smaller
amplitude of the fluctuations in initial time series. Switching-on
of the flickering stimulus causes the increasing of average value
of signals in healthy subject, and in patient's signals that
causes the more significant fluctuations.

Figs. 3, 4  demonstrate the plain phase projections of the phase
clouds of dynamic orthogonal variables $\{\textbf{W}_0^X,
\textbf{W}_i^Y\}$, where $i=0...3$. To represent the generalized
view of phase portraits of possible combinations of the dynamic
orthogonal variables $\{\textbf{W}_{0...3}^X,
\textbf{W}_{0...3}^Y\}$ we have chosen first four of them. On
phase portraits for healthy subject's brain signals (Fig. 3) two
phase areas similar on structure are distinctly manifested.
Smaller of them (it is shown by the arrow) corresponds to the
magnetic field dynamics when the flickering stimulus is switched
off.  Switching-on of stimulus causes the occurrence of transition
between the phase cloud areas that is connected with changes of
amplitude in initial signals (Fig. 1).

Other picture is observed in the phase portraits of brain signals
in patient with PSE (Fig. 4). First, phase clouds have bigger
sizes than those for the healthy subject. Secondly, they have
stratified structure concerning the central nucleus. It means,
that switching-on of stimulus does not cause the appreciable phase
changes in dynamics of the brain signals.

\subsection{Spectral cross-correlation properties of human neuromagnetic responses}

Power spectra of used cross-correlation functions for the studied
neuromagnetic signals allow to extract unique information about
the collective phenomena in brain functioning.

To demonstrate the general mechanisms of coordination in cerebral
cortex signals we present power spectra $\mu_0^X(\nu),
\mu_0^Y(\nu)$ of auto correlation functions for two signals and
power spectrum of respective cross-correlation function
$\mu_0^{XY}(\nu)$ in Fig. 5. These dependencies are derived for
neuromagnetic responses of one of the healthy subjects. Their
comparison distinctly detects the increasing of respective peaks
in CCF power spectrum (see, for example, peaks 1 and 2). Actually,
we reveal here a frequency-phase synchronization between the
cerebral cortex signals as a response to color flickering
influence. Thus analysis of the initial CCF power spectra allows
to detect synchronization of the brain signals and reveal its
frequencies.

Power spectra of initial CCF and respective memory functions for
the brain signals of healthy subject (Fig. 6) are characterized by
two peaks. The most significant of them is in the frequency 9.18
Hz, i.e. it is within the frequency range of normal brain rhythms.
These rhythms reflect the complex psycho-physiological processes
of brain activity. Therefore strong change of their characteristic
frequencies is the indicator of functioning abnormalities in human
brain and the central nervous system. The second peak has
frequency of 17.56 Hz. Besides memory functions power spectra
contain the additional peaks in higher frequencies. They reflect
specific features of the brain bioelectric activity. Note, that
their frequencies are multiple of 8.3 Hz, i.e. of frequencies of
the normal brain physiological rhythms. The analysis of CCF and
memory functions power spectra derived for combinations of control
group signals, shows, that all principal peaks are in the field of
low frequencies (up to 50 Hz), and the highest peaks have
frequency of 8 - 12 Hz. Thus, the frequency-phase synchronization
of brain signals in healthy people occurs in a low-frequency
range.

Just the opposite picture is observed in power spectra of
cross-correlation functions for brain signals of the patient with
PSE (Fig. 7). Peaks fill in all range of frequencies. Thus the
process with frequency of 49.33 Hz is dominating. Spectra in all
relaxation levels have the same kind.

Specific character in interaction of brain signals of the
patient with photosensitive epilepsy reflects abnormalities in
reaction to flickering stimulus. Considered power spectra allow to
take apart the development of abnormal high neuron collective
activity of spaced cerebral cortex areas, as an pathological
reaction to visual influences, resulting in epileptic seizure. At
the same time in healthy people we reveal the original protective
mechanism blocking the development of such reaction. At PSE the
suppression of this mechanism is reflected in domination of more
high-frequency processes on intensity comparable with the normal
physiological rhythms.

\subsection{Differentiation of intensity of the statistical memory at PSE}

To reveal the crucial role of statistical memory effects in
cerebral cortex activity at PSE we shall use the cross-correlation
information measure of memory $\varepsilon^{XY}_1 (0)$. At first
the non-Markovian parameter $\varepsilon$ was introduced in
statistical physics of condensed matter \cite{Shu1,Shu2}, however
later it proved itself in detecting the physical mechanisms of
abnormal functioning of live systems \cite{rmy1,rmy2,rmy3}.

Both non-Markovian parameter $\varepsilon$ and its generalization
$\varepsilon_1^{XY}(0)$ for cross-correlation analysis allow to
characterize the degree of statistical memory effects in
long-range component of discrete dynamics. If
$\varepsilon_1^{XY}(0) \gg 1$, then the Markovian components
dominate in dynamics of stochastic processes. In this case the
time of memory existence is too much shorter than relaxation time
of initial CCF. The weak (short) statistical memory is manifested.
Decrease of this information measure characterizes the expansion
of memory lifetime. $\varepsilon_1^{XY}(0) \sim 1$ means that the
processes are characterized by long-range (strong) statistical
memory. In this case the time of memory existence is covariant
with the relaxation time of initial CCF. When
$\varepsilon_1^{XY}(0)
> 1$ we can consider the studied processes to be quasi-Markovian
with moderate (intermediate on time existence) statistical memory.
Thus, the introduced quantitative criterium allows to parameterize
the memory effects intensity and the velocity of loss of
relaxation processes.

In Figs. 8, 9 we have presented the frequency spectra of
information measures of memory $\varepsilon_i^{51-56}(\nu)$, where
$i=1...3$, for neuromagnetic responses of spaced SQUIDs No. 51,
No. 56. Statistical memory intensity in time intervals about
length of initial series is determined by the low-frequency areas,
and the high-range frequencies correspond to smaller time scales.
To study the general manifestations of statistical memory effects
in initial time series we shall consider only the parameter
$\varepsilon_1^{51-56}(0)$. For healthy subject the value of this
parameter is $9.15$. It means that the statistical memory effects
are decreased in scales of time series length at the same time.
For patient the parameter $\varepsilon_1^{51-56}(0)$ value is
$1.61$. Frequency dependences of $\varepsilon_{2,3}^{51-56}(\nu)$
in case of healthy subject (see Figs. 8b, c) are characterized by
the significant peaks in low- and high-frequency ranges against
the background of general oscillation structure with aliquot
frequencies. At PSE (Figs. 9b, c) such structure of information
measures spectra $\varepsilon_{2,3}^{51-56}(\nu)$ is depressed.

To study the general character of statistical memory manifestation
in the dynamics of neuromagnetic responses of spaced cerebral
cortex areas we derived the ratios of averaged values of the
information measure $\varepsilon_1^{XY}(0)$ for control group and
respective value for patient with PSE. These ratios are derived
for all combinations of considered SQUIDs and are presented in
Table 1. Analysis of the presented values allows to divide the
SQUIDs combinations into three groups:

\begin{table}[h]
 \centering
 \footnotesize
  \caption{Ratio of averaged value of information measure
  $\varepsilon_1^{XY}(0)$ for control group and respective value
  for patient with PSE. Considered SQUIDs are located in occipital
  (No. 51, 52, 53), left-temporal (No 30, 34) and right-temporal
  (No. 56, 57) areas of the cerebral cortex}
  \scriptsize\begin{center}
  \begin{tabular}{|p{2.1cm}|p{2.7cm}|p{2.7cm}|p{2.7cm}|p{2.7cm}|}
    \hline
    ~~\quad SQUID&~~~~~~\qquad30&~~~~~~\qquad34&~~~~~~\qquad56&~~~~~~\qquad57\\\hline
    ~~~\qquad51&~~~~~\qquad4.22&~~~~~\qquad24.10&~~~~~\qquad6.03&~~~~~\qquad20.48\\\hline
    ~~~\qquad52&~~~~~\qquad0.92&~~~~~\qquad1.45&~~~~~\qquad0.72&~~~~~\qquad2.42\\\hline
    ~~~\qquad53&~~~~~\qquad3.69&~~~~~\qquad37.35&~~~~~\qquad4.28&~~~~~\qquad3.40\\\hline
  \end{tabular}
\end{center}
\normalsize
\end{table}

\begin{enumerate}

\item SQUID combinations: 52-30, 52-34, 52-56, 52-57. The
differences in information measures $\varepsilon_1^{XY}(0)$ are
less than of equal to 2.5 times. At PSE the coupling of signals,
generated by these cerebral cortex areas, considerably does not
change.

\item SQUID combinations: 51-30, \textbf{51-56}, 53-30, 53-56,
53-57. At PSE the mutual dynamics of neuromagnetic responses,
generated by these areas is characterized by the significant
increase of memory effects, that may be considered as the
diagnostic indicator of pathological changes.

\item Finally the mutual behavior of signals, registered by SQUIDS
51-34, 51-57, 53-34 is characterized by the most dramatic (in 20
-- 37.5 times) increase of statistical memory intensity at PSE.
Coupling of these areas changes most strongly.

\end{enumerate}

Actually, parameter $\varepsilon_1^{XY}(0)$ allows to quantitative
differ the manifestation of statistical memory effects in mutual
dynamics of neuromagnetic responses in patient with PSE. Thus the
possibility of revealing the peculiar zones of anomalous
collective neuron excitation (as an example, SQUID No. 51) at PSE,
at which the depression of brain regulator functions, as a
response to flickering color stimuli occurs. Note that earlier in
paper \cite{rmy3} some conclusions have been made about the PSE's
abnormalities of MEG signal co-ordinations at influence of
flickering stimuli.

\section{Conclusion. The role of cross-correlations and collective effects in complex systems}

Time signals generated by a complex system contain the
unique information about its organization and character of
couplings between its constituent systems. Therefore a representation of this
information as a set of variables, i.e. parametrization of complex
systems dynamics, is crucial to understand the complex system. To analyze the quantitative
and qualitative parameters of the mutual dynamics of complex
system, we here use the cross-correlation functions.

To study the mechanisms of collective phenomena in dynamics
of cerebral cortex, we analyze MEG signals from healthy healthy subjects and a patient with
PSE while viewing chromatic flickering stimuli, we have used generalization of memory functions formalism in case of cross-correlation analysis.

Analysis of cross-correlations and collective effects in the
mutual dynamics of MEG responses reveals a protective mechanism present in healthy brain, which acts
against adverse external perturbations and manifests in certain regime of
frequency-phase synchronization. In healthy subjects this mechanism blokes the developing of epileptic reaction to the influence of flickering light. In this case power spectra of the initial cross-correlation function and memory functions demonstrate a clear dominance of the low-frequency processes in coupling of MEG responses, respective to the
normal brain rhythms. At the same time the
high-frequency processes become dominant in the MEG responses brain signals of the
patient with PSE. Therefore it is reputed that the substitution of low-frequency synchronization of the MEG signals by high-frequency synchronization at flickering light influence is an reliable indicator of PSE.

The proposed analysis has offered not only a new insight into the
character of the pathological changes in patient's
brain responses but also reveals a high degree of individuality
in neuromagnetic brain responses.

\section{Acknowledgments}
This work is supported by Russian Foundation for Basic Research
(Grant No. 08-02-00123-a). We thank K. Watanabe and S. Shimojo for the experimental data.
We also thank Anatolii V. Mokshin and Ramil M. Kusnutdinoff for their help with numerical
computations and Elvira R. Nigmatzyanova for technical assistance.

\newpage
\begin{figure}
\includegraphics[height=10cm,angle=0]{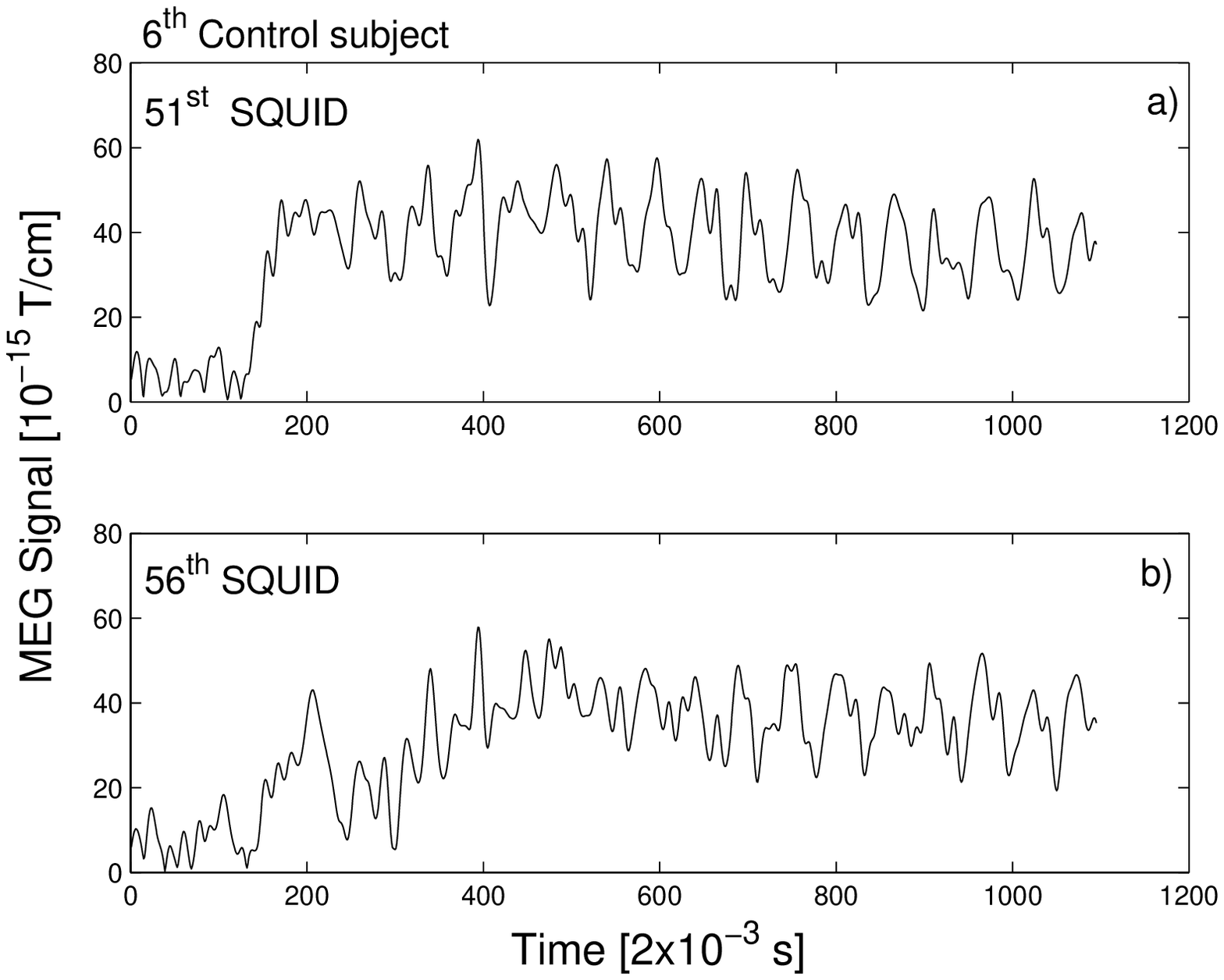}
\caption{Induced neuromagnetic responses to the red-blue
flickering stimulus from the occipital (51-st SQUID) and the
right-temporal (56-th SQUID) cerebral cortex areas of healthy
subject. The large-scale fluctuations against the quasi-periodic
background are connected with manifesting the physiological
rhythms of brain bioelectric activity}
\end{figure}

\newpage
\begin{figure}
\includegraphics[height=10cm,angle=0]{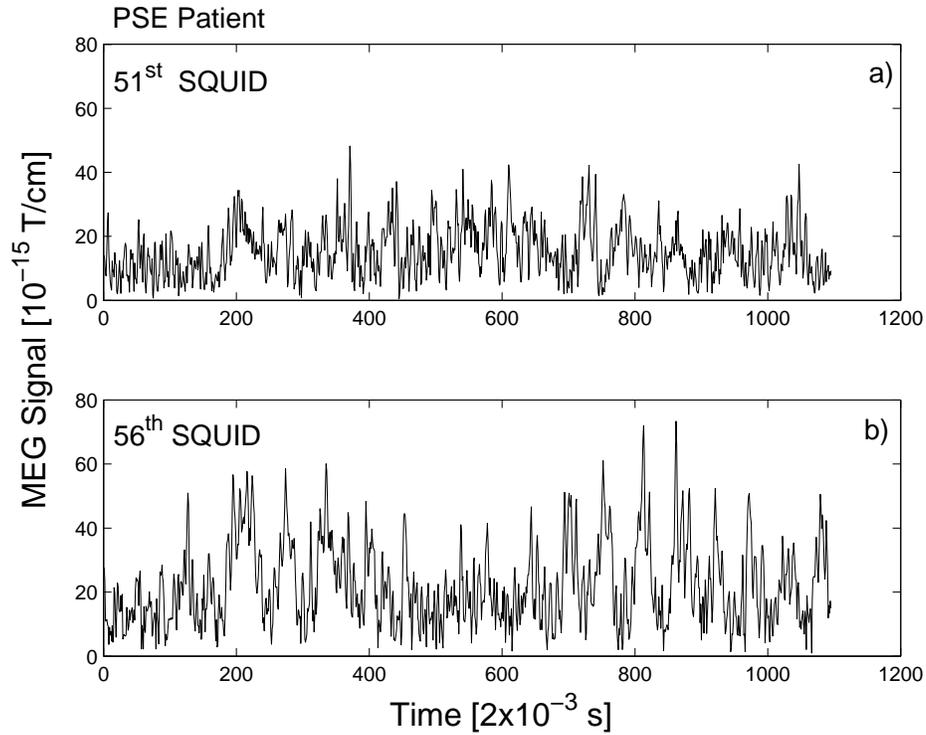}
\caption{Magnetoelectric signals, generated by occipital (51-st
SQUID) and right-temporal (56-th SQUID) cerebral cortex areas of
patient with PSE under the influence of red-blue flickering
stimulus. Signal structure is determined by the superposition of
the small-scale fluctuations against the quasi-oscillatory
background}
\end{figure}

\newpage
\begin{figure}
\includegraphics[height=10cm,angle=0]{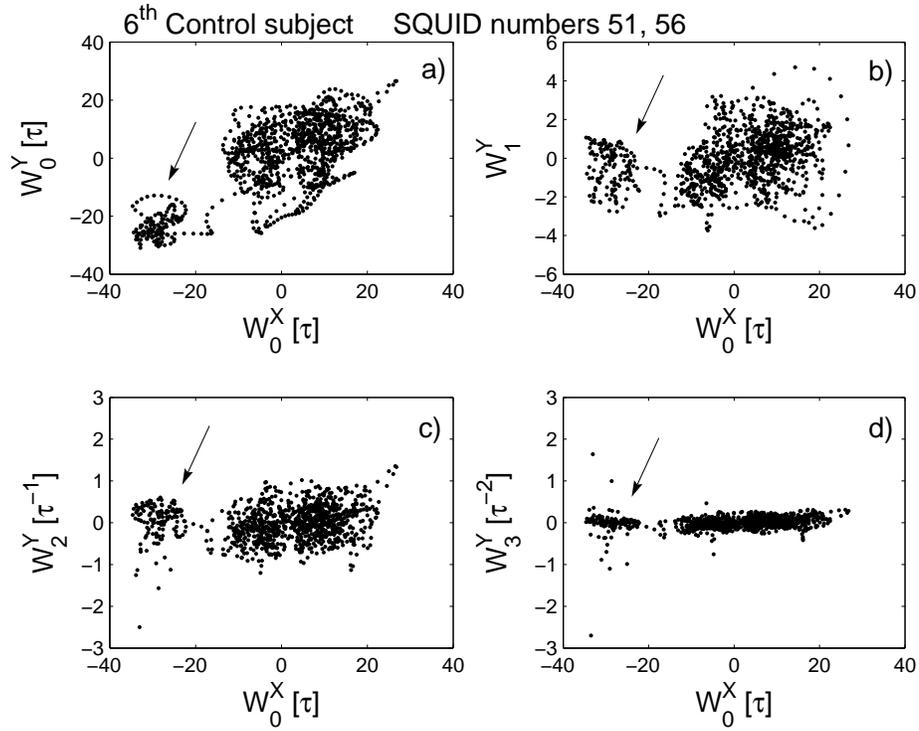}
\caption{Typical structure of plane portraits of the dynamic
orthogonal variables $\textbf{W}_i^Y=f(\textbf{W}_0^X)$, where
$i=0...3$, for MEG signals of healthy subject. Areas of phase
clouds, indicated by the arrow, correspond to neuromagnetic
responses in absence of external flickering stimulus}
\end{figure}

\newpage
\begin{figure}
\includegraphics[height=10cm,angle=0]{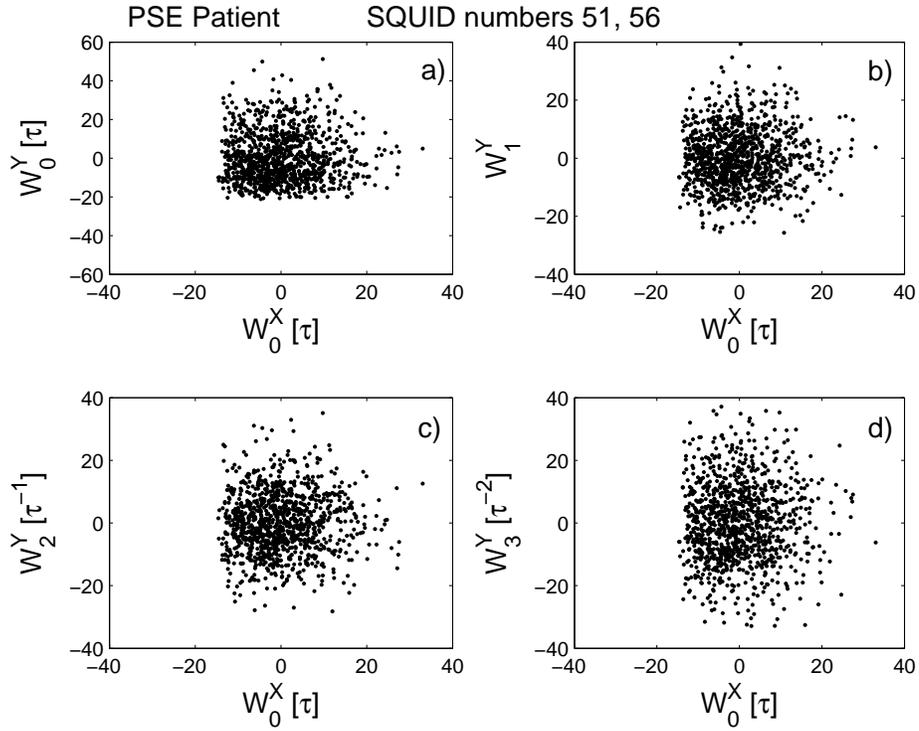}
\caption{Plane projections of phase portraits of the dynamic
orthogonal variables $\{\textbf{W}_0^X, \textbf{W}_i^Y\}$, where
$i=0...3$, for the induced cerebral cortex signals from patient
with photosensitive epilepsy. The apparent increasing of scales
for phase clouds is revealed}
\end{figure}

\newpage
\begin{figure}
\includegraphics[height=10cm,angle=0]{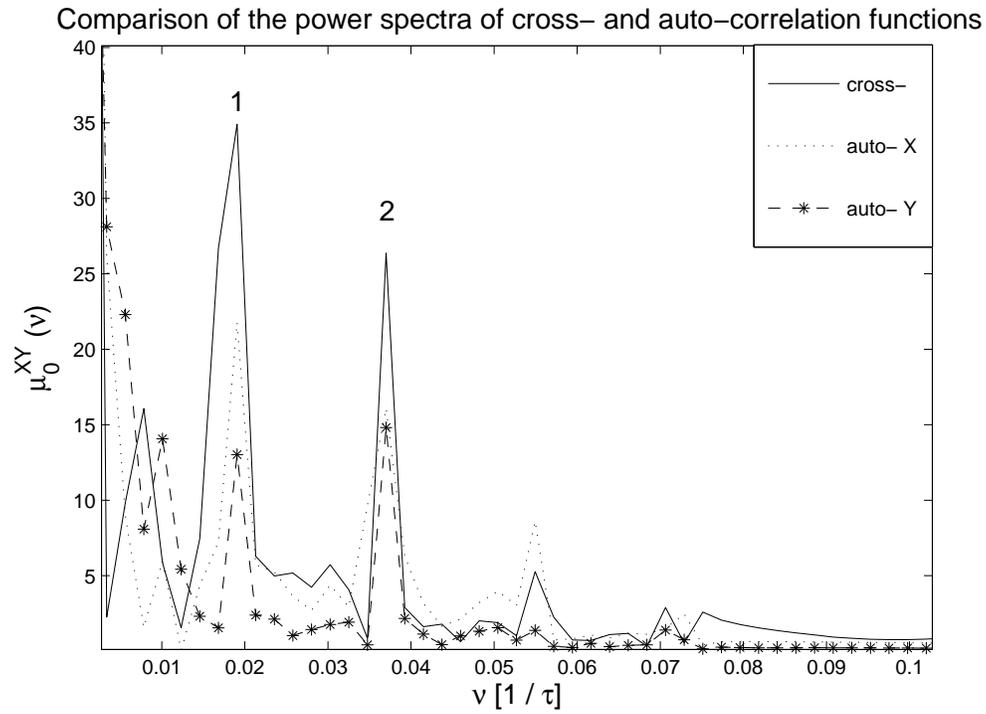}
\caption{Power spectra of auto-correlation functions for signals
$X$ and $Y$ in comparison with power spectra of cross-correlation
function $c(t)$. Frequency synchronization of these signals results
in increasing of respective peaks on power spectrum of cross-correlation function.}
\end{figure}

\newpage
\begin{figure}
\includegraphics[height=10cm,angle=0]{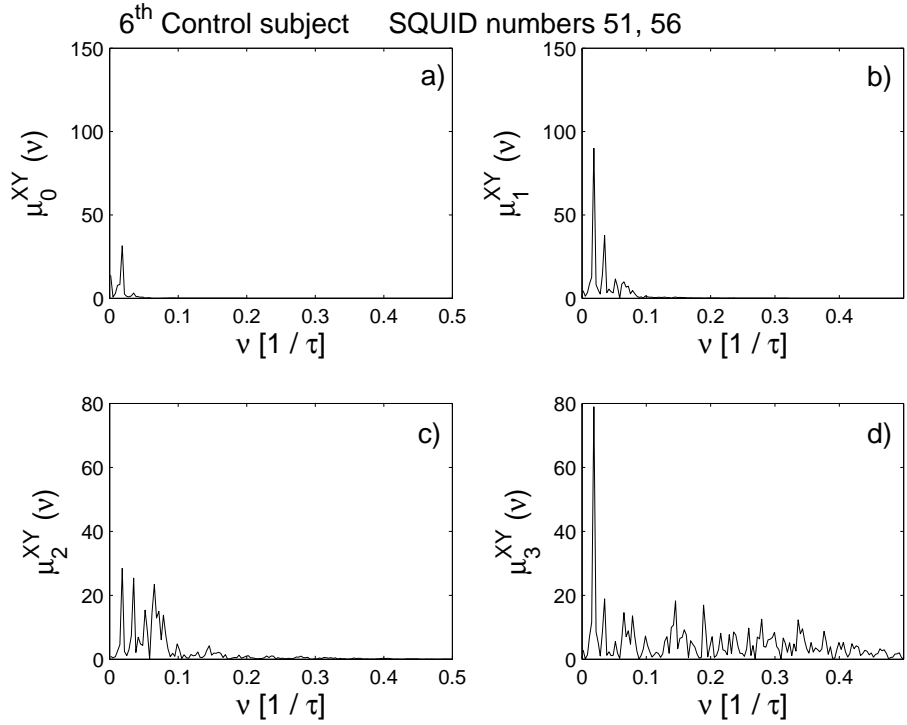}
\caption{Typical view of power spectra of the initial cross
correlation function and memory functions for the mutual dynamics
of neuromagnetic responses of healthy subject. Peaks, which
characterize the periodic features of brain neuromagnetic
activity, are in a low-frequency range of physiological rhythms}
\end{figure}

\newpage
\begin{figure}
\includegraphics[height=10cm,angle=0]{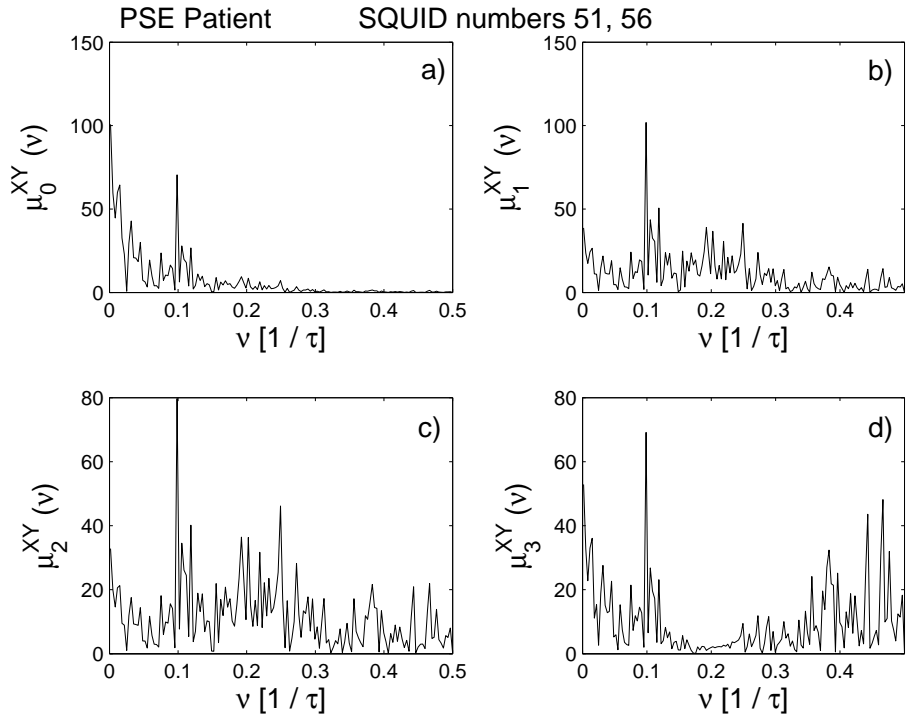}
\caption{Power spectra of the initial cross correlation function
and memory functions for induced MEG signals from patient with
photosensitive epilepsy. As it is seen from the specific
low-frequency dynamics in healthy subject, the suppression of
normal physiological rhythms by high frequency processes is
observed in this case. The frequency synchronization abnormality
can be interpreted as a diagnostic indicator of PSE pathological
manifestation.}
\end{figure}

\newpage
\begin{figure}
\includegraphics[height=6.5cm,width=16cm,angle=0]{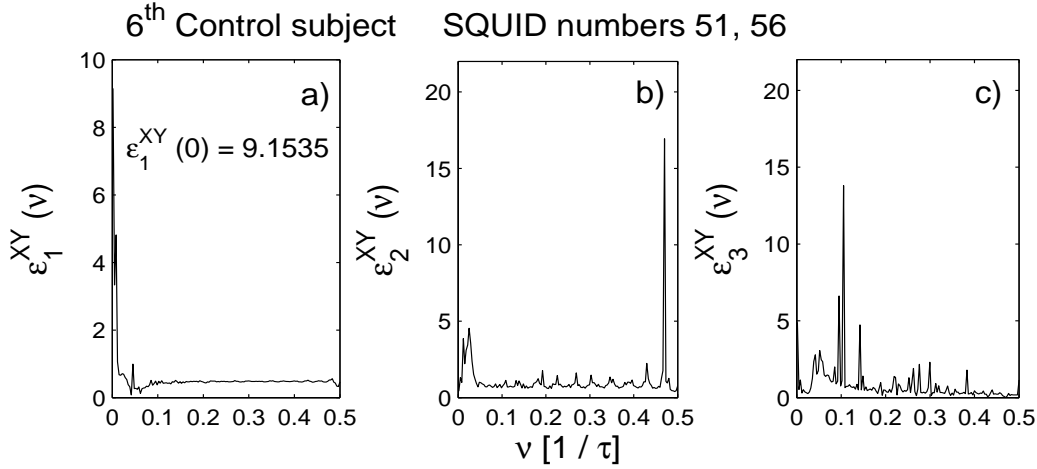}
\caption{Frequency dependences of the information measures of
memory $\varepsilon_i^{XY}(\nu)$, where $i=1...3$, for signals,
registered from 51-st and 56-th SQUIDs. Mutual dynamics of these
signals is characterized by quasi-Markovity and sufficient weak
intensity of statistical memory. Oscillatory type of spectra of
the information measure $\varepsilon_{2,3}^{XY}(\nu)$ is connected
with manifesting the brain rhythms in MEG signals of healthy
subject}
\end{figure}

\newpage
\begin{figure}
\includegraphics[height=6.5cm,width=16cm,angle=0]{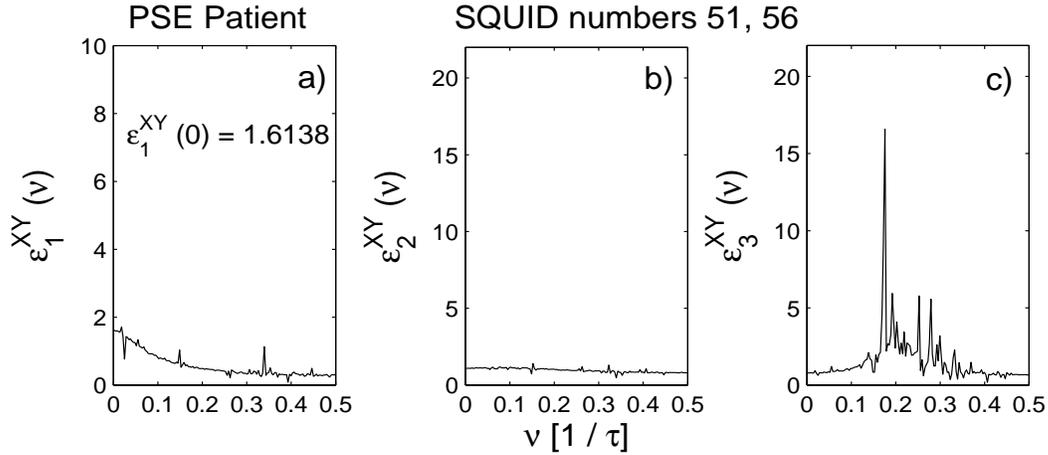}
\caption{Frequency spectra  $\varepsilon_i^{XY}(\nu)$, where
$i=1...3$, for neuromagnetic responses from cerebral cortex of
patient with PSE. The typical alteration of the intensity of
statistical memory in dynamics of patient's neuromagnetic
responses to external light influence in comparison with control
group is observed. According to this fact the effects of
statistical memory play a crucial role in pathological changes in
case of PSE}
\end{figure}

\end{document}